\documentclass[prd,showpacs,amsmath,amssymb,floatfix,nofootinbib]{revtex4}
\usepackage{graphicx,color,dcolumn,,booktabs}
\usepackage{longtable,lscape}

\def\lrpartial{\buildrel\leftrightarrow\over\partial}
\begin{document}
\title{Long-distant contribution and $\chi_{c1}$ radiative decays to light vector meson}
\author{Dian-Yong Chen$^{1,2}$}
\author{Yu-Bing Dong$^{4,5}$}\author{Xiang Liu$^{1,3}$\footnote{Corresponding author}}\email{xiangliu@lzu.edu.cn}
\affiliation{$^1$Research Center for Hadron and CSR Physics, Lanzhou
University $\&$ Institute of Modern Physics of CAS, Lanzhou 730000,
China\\
$^2$Institute of Modern Physics, Chinese Academy of Sciences,
Lanzhou 730000, China\\
$^3$School of Physical Science and Technology, Lanzhou University,
Lanzhou 730000, China\\
$^4$ Institute of High Energy Physics of CAS, Beijing, 100049,
China\\
$^5$ Theoretical Physics Center for Science Facilities (TPCSF), CAS,
Beijing 100049, China}
\date{\today}
\begin{abstract}

The discrepancy between the PQCD calculation and the CLEO data for
$\chi_{c1}\to \gamma V$ ($V=\rho^0,\,\omega,\,\phi$) stimulates our
interest in exploring other mechanisms of $\chi_{c1}$ decay. In this
work, we apply an important non-perturbative QCD effect, i.e.,
hadronic loop mechanism, to study $\chi_{c1}\to \gamma V$ radiative
decay. Our numerical result shows that the theoretical results
including the hadronic loop contribution and the PQCD calculation of
$\chi_{c1}\to \gamma V$ are consistent with the corresponding CLEO
data of $\chi_{c1}\to \gamma V$. We expect further experimental
measurement of $\chi_{c1}\to \gamma V$, which will be
helpful to test the hadronic loop effect on $\chi_{c1}$ decay.

\end{abstract}
\pacs{14.40.Pq, 11.30.Hv, 12.39.Fe, 12.39.Hg}

\maketitle

\section{Introduction}\label{sec1}

In the past three decades, a series of the observations of S-wave,
P-wave and D-wave charmonia make charmonium family abundant.
Nowadays, charm physics is still an intriguing research field with
challenges and opportunities \cite{Li:2008ey}. Especially, the study
of charmonium may provide valuable information on non-perturbative
QCD effects.

As an important and effective approach to deeply learn the underlying
properties of charmonium, charmonium decay is an extensively focused
research topic. Among the observed charmonium states, $J/\psi$ and
$\psi(2S)$ are of abundant experimental information of decay just
listed in Particle Data Group (PDG) \cite{Amsler:2008zzb}. However,
the experimental measurement relevant to the decay of P-wave
charmonium is far less than that of $J/\psi$ and $\psi(2S)$. Thus,
more experimental and theoretical explorations of P-wave charmonium
decay are becoming active, especially with the running of BES-III.

Recently, the BES-III Collaboration announced its observations of
$\chi_{cJ} (J=0,1,2)$ decaying into two light vector mesons
\cite{zhangJ}. Among these decay modes of $\chi_{cJ}$ to two light
vector mesons, the OZI suppressed processes $\chi_{c1} \to \omega
\omega,\ \phi \phi$ and the double-OZI suppressed process
$\chi_{c1}\to \omega \phi$ were firstly observed. In order to
explain the evasion of the helicity selection rule in these
processes, the hadronic loop effect, an important non-perturbative
effect relevant to the decay of charmonia \cite{Lipkin:1986av,Liu:2006dq,Li:2007xr,Liu:2009dr,Liu:2009iw} and molecular system \cite{He:2006is,Liu:2007qi,Liu:2007fe,Liu:2007ez,Liu:2008yy,Dong:2009tg,Faessler:2007gv,Faessler:2007us,Faessler:2008vc,Dong:2008gb,Dong:2009yp,Branz:2009yt,Dong:2010gu}, is introduced in
Refs. \cite{Chen:2010rd,Liu:2010rd,Liu:2006df}, which also indicate
that applying the hadronic loop mechanism to other $\chi_{cJ}$
decays will be helpful to further test the hadronic loop effect on
$\chi_{cJ}$ decay.

In Ref. \cite{Gao:2006bc}, the radiative decays of charmonia
$J/\psi$ and $\chi_{cJ}$ into light meson are studied by the
perturbative QCD (PQCD) approach, where a complete numerical
calculation for the quark-gluon loop diagrams was performed. The
obtained theoretical results for $J/\psi\to \gamma \eta,\,\gamma
\eta^\prime$ can well reproduce the experimental data. Furthermore,
the branching ratios of $\chi_{cJ} \to \gamma \rho^0,\,
\gamma\omega,\,\gamma \phi$ were predicted, which are
$\mathcal{B}(\chi_{c1} \to \gamma \rho^0)= 1.4 \times 10^{-5}$,
$\mathcal{B}(\chi_{c1} \to \gamma \omega) =1.6 \times 10^{-6}$ and
$\mathcal{B}(\chi_{c1} \to \gamma \phi) =3.6 \times 10^{-6}$.

In 2008, the radiative decays of $\chi_{cJ}$ were first measured by
the CLEO Collaboration using a total of $2.74 \times 10^7$ decays of
the $\psi(2S)$ collected with the CLEO-c detector
\cite{Bennett:2008aj}. The reported results are $\mathcal{B}
(\chi_{c1} \to \gamma \rho^0) =243 \pm 19 \pm 22 \times 10^{-6} $,
$\mathcal{B} (\chi_{c1} \to \gamma \omega) =85 \pm 15 \pm 12 \times
10^{-6}$ and $\mathcal{B} (\chi_{c1} \to \gamma \phi) =12.8 \pm 7.6
\pm 1.5 \times 10^{-6}$. One notices that the experimental results
of the branching ratios of $\chi_{c1} \to \gamma \rho^0,\, \gamma
\omega,\, \gamma \phi$ are an order magnitude larger than the
corresponding theoretical predictions \cite{Gao:2006bc}.

The above difference between the experimental measurement from CLEO
and the theoretical result calculated in PQCD shows that there
should exist the extra effect on the $\chi_{c1}$ radiative decays
into a light vector meson, which stimulates our interest in
exploring the underlying mechanism to resolve this large discrepancy
between the theoretical prediction by PQCD \cite{Gao:2006bc} and the
experimental measurement by CLEO \cite{Bennett:2008aj}.

As indicated in Refs. \cite{Chen:2010rd,Liu:2010rd}, the hadronic
loop effect can explain the experimental observation of the OZI
suppressed processes $\chi_{c1} \to \omega \omega,\ \phi \phi$ and
the double-OZI suppressed process $\chi_{c1}\to \omega \phi$ well.
In this work, we extend the hadronic loop effect to $\chi_{c1}\to
\gamma V$ ($V=\rho^0,\,\omega,\,\phi$) process to answer whether
large discrepancy between the theoretical calculation and the
experimental measurement of $\chi_{c1}\to \gamma V$ can be
alleviated. $\chi_{c1}\to \gamma V$ is similar to $\chi_{c1}\to VV$
since both processes occur via the intermediate charmed mesons under
considering the hadronic loop effect. What is more important of
this work is to offer an effective approach to test the proposed
hadronic loop effect applied to explain $\chi_{c1}\to VV$ decay
processes.

The paper is organized as following. After the introduction, we
present the formula of the hadronic loop contributions to $\chi_{c1}$
radiative decays to a light vector meson, which includes the effective
Lagrangian employed in this work and the decay amplitudes. In Sec. \ref{sec3},
the numerical results of $\chi_{c1} \to \gamma \rho^{0},\ \gamma \omega,\ \gamma
\phi$ are given. The last section is the discussion and conclusion.

\section{Hadronic loop effect on $\chi_{c1}\to \gamma V$ }\label{sec2}

\begin{center}
\begin{figure}[htb]
\begin{tabular}{cccc}
Quark level&Hadron level\\
\raisebox{5ex}{\scalebox{0.55}{\includegraphics{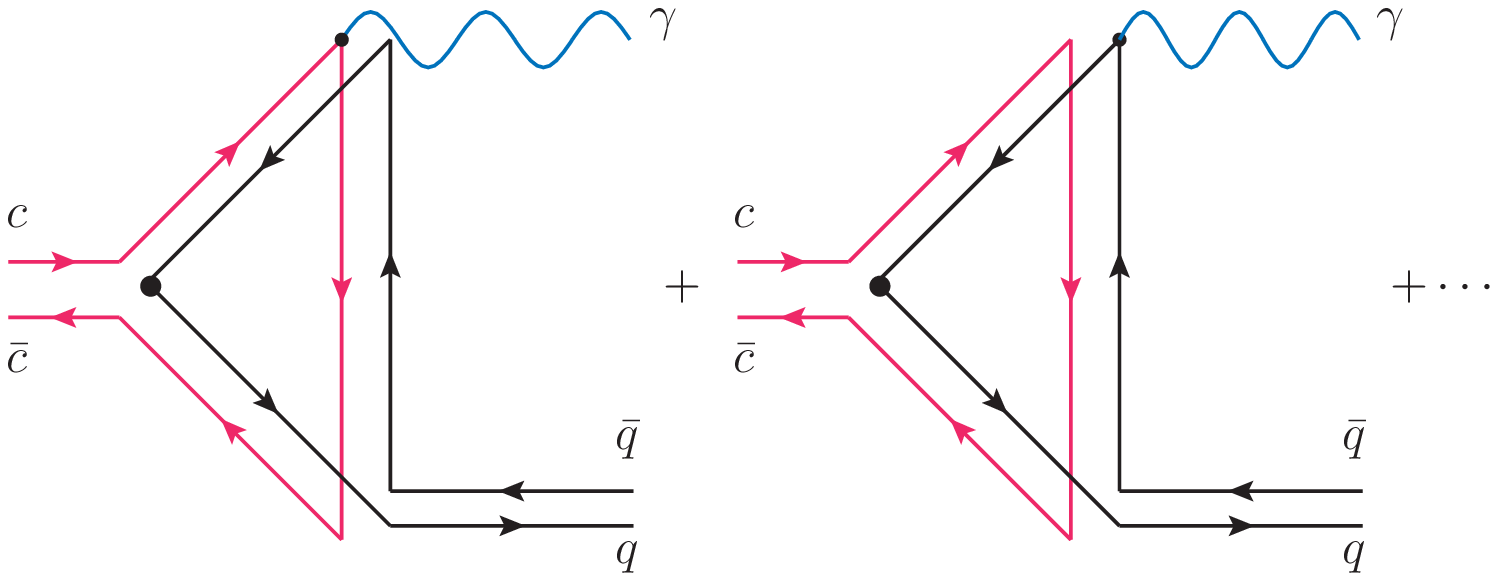}}}&
\scalebox{0.55}{\includegraphics{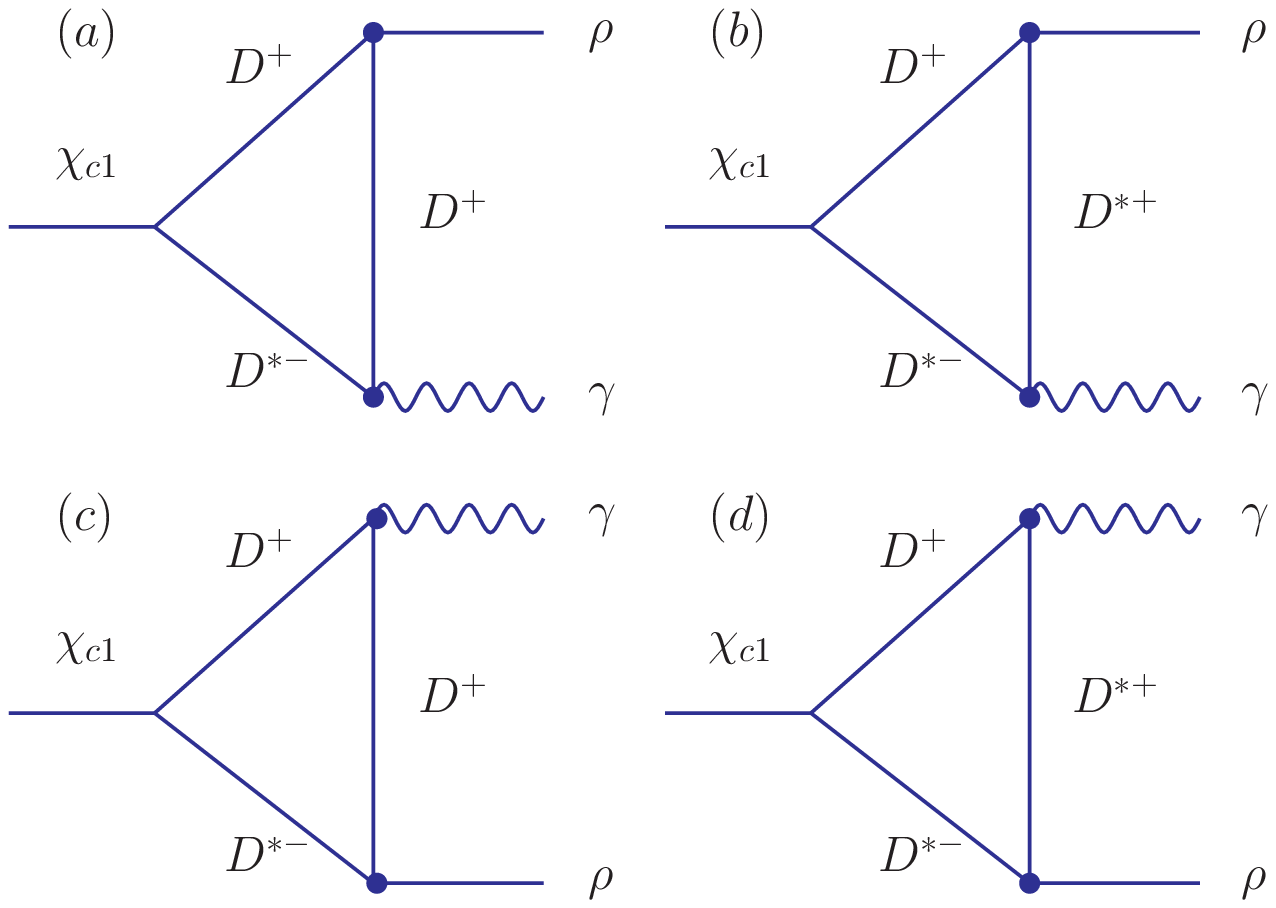}}\\
\end{tabular}
\caption{The quark level typical diagrams (the first column) describing the hadronic loop effect on $\chi_{c1}\to \gamma V$ and the hadron level schematic diagram corresponding to $\chi_{c1}\to \gamma\rho^0$ (the second column). The red and black lines denote the charm quark and light quark. The photon emits from charm quark line or light quark line. By the charge conjugate transformation $D^{(*)+}\rightleftharpoons D^{(*)0}$ and $D^{(*)-}\rightleftharpoons \bar D^{(*)0}$, the rest two diagrams of $\chi_{c1}\to D\bar{D}^*+h.c.\to\gamma\rho^0$ can be obtained by diagrams (a) and (d).  \label{Fig:gpept-1}}
\end{figure}
\end{center}

As indicated in Ref. \cite{Chen:2010rd}, the nonperturbative QCD mechanism, i.e., hadronic loop effect, may play a crucial role in understanding $\chi_{c1}$ decay. In Table. \ref{Fig:gpept-1}, the typical diagram depicting the hadronic loop effect on $\chi_{c1}\to \gamma V$ at quark level is given, which is different from the quark-gluon loop diagrams depicting $\chi_{c1}\to
\gamma V$ at PQCD approach in Ref. \cite{Gao:2006bc}. The transition element of $\chi_{c1}\to \gamma V$ can be expressed as
\begin{eqnarray}
\mathcal{M}[\chi_{c1}\to \gamma V]=\sum_i \langle \gamma V|\mathcal{H}^{(2)}|i\rangle \langle i |\mathcal{H}^{(1)}|\chi_{c1}\rangle,
\end{eqnarray}
which reflects the intermediate state contribution to $\chi_{c1}\to \gamma V$. Here, $\mathcal{H}^{(1)}$
represents the interaction of $\chi_{c1}$ and $D\bar{D}^*+h.c.$ and $\mathcal{H}^{(2)}$ describes interaction $D\bar{D}^*+h.c. \to \gamma V$ by exchanging an appropriate charmed meson.

\renewcommand{\arraystretch}{1.50}

With $\chi_{c1}\to \gamma \rho^0$ as an example, we list the schematic diagrams at hadron level in Table \ref{Fig:gpept-1}, where $\chi_{c1}$ first dissolves into two virtual charmed mesons which is originated from the coupled channel effect. Then these two virtual charmed mesons $D\bar{D}^*+c.c.$ turn into a photon and $\rho^0$ meson by exchanging the charmed meson. Due to the mass of $\chi_{c1}$ being
lower than the threshold of $D\bar{D}^*$, the charmed mesons in the loop
are off-shell.

In the following, we still use $\chi_{c1}\to \gamma \rho^0$ as an example to illustrate the relevant calculation of hadronic loop diagrams listed in Fig. \ref{Fig:gpept-1}, where the effective Lagrangian approach is applied to write out the decay amplitude throughout this work.

The Lagrangian for $\chi_{c1} \mathcal{D}\mathcal{D}^{*}$ coupling
reads \cite{Colangelo:2003sa}
\begin{eqnarray}
\mathcal{L}_{\chi_{1}\mathcal{DD}^{*}} = i g_{\chi_{c1} \mathcal{D}
\mathcal{D}^{*}} \chi_{c1} \cdot \mathcal{D}^{*\dagger}_{i}
\mathcal{D}^{i} + h.c.
 \label{Eq:chic1DD}%
\end{eqnarray}
The effective Lagrangians responsible for $\mathcal{D}^{(*)} \mathcal{D}^{(*)} V$ interactions are
\begin{eqnarray}
\mathcal{L}_{\mathcal{D}^{(*)} \mathcal{D}^{(*)} \mathcal{V}}
&=&-ig_{\mathcal{DDV}} \mathcal{D}_i^\dagger \lrpartial_{\!\mu}
\mathcal{D}^j (\mathcal{V}^\mu)^i_j -2f_{\mathcal{D}^* \mathcal{D}
\mathcal{V}} \epsilon_{\mu\nu\alpha\beta} (\partial^\mu
\mathcal{V}^\nu)^i_j (\mathcal{D}_i^\dagger\lrpartial{}^{\!\alpha}
\mathcal{D}^{*\beta j}
-\mathcal{D}_i^{*\beta\dagger}\lrpartial{}{\!^\alpha}
\mathcal{D}^j)\nonumber\\
&&+ ig_{\mathcal{D}^*\mathcal{D}^*\mathcal{V}}
\mathcal{D}^{*\nu\dagger}_i \lrpartial_{\!\mu}
\mathcal{D}^{*j}_\nu(\mathcal{V}^\mu)^i_j
+4if_{\mathcal{D}^*\mathcal{D}^*\mathcal{V}}
\mathcal{D}^{*\dagger}_{i\mu}(\partial^\mu
\mathcal{V}^\nu-\partial^\nu \mathcal{V}^\mu)^i_j
\mathcal{D}^{*j}_\nu,
 \label{Eq:DDV}
\end{eqnarray}
where $\mathcal{D}^{(*)}=(D^{(*)0},D^{(*)+},D_s^{(*)+})$ and $A
\lrpartial_{\mu} B =A (\partial_{\mu} B) - (\partial_{\mu} A) B $.
The matrix of the nonet vector mesons $\mathcal{V}$ is defined as
\begin{eqnarray}
\mathcal{V}=
 \left(
 \begin{array}{ccc}
\frac{1}{\sqrt{2}} (\rho^{0}+ \omega) & \rho^{+} & K^{*+}\\
\rho^{-} & \frac{1}{\sqrt{2}}(-\rho^0 +\omega) &  K^{*0}\\
 K^{*-} & \bar{K}^{*0} & \phi
 \end{array}
 \right).
\end{eqnarray}
We need to specify that the Lagrangians in Eq. (\ref{Eq:DDV}) is just the first term in an infinite series of terms that represents the hadronic representation of the QCD Lagrangian, which is restored from the Lagrangian constructed in the chiral and heavy quark limits \cite{Cheng:1992xi, Yan:1992gz, Wise:1992hn, Burdman:1992gh,Casalbuoni:1996pg}. Thus, we adopt usual $i/(k^2-m^2)$ and $i(-g^{\mu\nu}+k^\mu k^\nu/m^2)/(k^2-m^2)$ propagators when writing out the decay amplitude of $\chi_{c1}\to D\bar{D}^*+h.c.\to \gamma \rho^0$. The relevant coupling constant will presented in the next section.

The Lagrangians for $\gamma DD$ and $\gamma D^{*}D^{*}$ interactions
can be obtained from the Lagrangian for free scalar and massive
vector fields by the minimal substitution $\partial^{\mu} \to
\partial^{\mu}+ie A^{\mu}$, which are \cite{Dong:2009uf}
\begin{eqnarray}
\mathcal{L}_{\mathcal{DD}\gamma} = i e A_{\mu} D^{-} \lrpartial^{\mu}
D^{+}+ i e A_{\mu} D_s^{-} \lrpartial^{\mu}
D_s^{+},
 \label{Eq:DDgamma}
\end{eqnarray}
\begin{eqnarray}
\mathcal{L}_{\mathcal{D}^{*} \mathcal{D}^{*} \gamma} &=& i e A_{\mu} \left\{ g^{\alpha
\beta} D_{\alpha}^{*-} \lrpartial^{\mu} D_{\beta}^{*+} + g^{\mu
\beta} D_{\alpha}^{*-} \partial^{\alpha} D_{\beta}^{*+} - g^{\mu
\alpha }
\partial^{\beta} D_{\alpha}^{*-}  D_{\beta}^{*+}\right\}\nonumber\\
&&+i e A_{\mu} \left\{ g^{\alpha
\beta} D_{s\alpha}^{*-} \lrpartial^{\mu} D_{s\beta}^{*+} + g^{\mu
\beta} D_{s\alpha}^{*-} \partial^{\alpha} D_{s\beta}^{*+} - g^{\mu
\alpha }
\partial^{\beta} D_{s\alpha}^{*-}  D_{s\beta}^{*+}\right\},
 \label{Eq:DsDsgamma}
\end{eqnarray}
respectively. Here, the electromagnetic interactions of $D^0D^0\gamma$ and $D^{*0}D^{*0}\gamma$ does not exist.

The Lagrangian describing the electromagnetic
${D}^{*}{D} \gamma$ vertex is \cite{Dong:2009uf}
\begin{eqnarray}
\mathcal{L}_{\mathcal{D}^{*} \mathcal{D} \gamma} = \bigg\{\frac{e}{4}
g_{{D^{*+}D^+}\gamma} \varepsilon^{\mu \nu \alpha \beta} F_{\mu
\nu} {D}^{*+}_{\alpha \beta} {D}^- +\frac{e}{4}
g_{{D^{*0}{D}^0}\gamma} \varepsilon^{\mu \nu \alpha \beta} F_{\mu
\nu} \mathcal{D}^{*0}_{\alpha \beta} \bar{{D}}^0\bigg\}+ h.c \ \ .
 \label{Eq:DsDgamma}
\end{eqnarray}
where $F_{\mu \nu} =\partial_{\mu} A_{\nu} - \partial_{\nu} A_{\mu}$
and $D^{*0,+}_{\mu \nu} = \partial_{\mu} D^{*0,+}_{\nu} -
\partial_{\nu} D^{*0,+}_{\mu}$ is the stress tensor of the vector
charmed meson. To some extent, the Lorentz structure just shown in Eq. (\ref{Eq:DsDgamma}) is same as
that in Ref. \cite{Colangelo:1994jc,Zhu:1996sr}. In Eq. (\ref{Eq:DsDgamma}), parameters $g_{{D^{*+}D^+}\gamma}$
and $g_{{D^{*0}{D}^0}\gamma}$ are introduced to get consistent results with experimental measurements of $D^{*+}\to D^+\gamma$ and $D^{*0}\to D^0\gamma$. The theoretical decay widths of $D^{*+}\to D^+\gamma$ and $D^{*0}\to D^0\gamma$ are
\begin{eqnarray}
\Gamma(D^{* +} \to D^{+} \gamma)&=& \frac{\alpha}{24} g_{D^{*+} D^{+}
\gamma}^2 m_{D^{* +}}^3 \bigg(1- \frac{m^2_{D^{+}}}{m^2_{D^{*+}}}\bigg),\\
\Gamma(D^{* 0} \to D^{0} \gamma) &=& \frac{\alpha}{24} g_{D^{*0} D^{0}
\gamma}^2 m_{D^{* 0}}^3 \bigg(1- \frac{m^2_{D^{0}}}{m^2_{D^{*0}}}\bigg).
\end{eqnarray}
According to the experimental widths $\Gamma(D^{*+}\to D^+\gamma)=1.54$ keV \cite{Amsler:2008zzb} and $\Gamma(D^{*0}\to D^0\gamma)=26.04$ keV \cite{Dong:2008gb,Amsler:2008zzb}, the coupling constant $g_{D^{*} D \gamma}$ is
fixed as
\begin{eqnarray}
|g_{D^{*+} D^{+} \gamma}|=0.5 \,\mathrm{GeV}^{-1}, \ \
|g_{D^{*0}D^{0}\gamma}| = 2.0\, \mathrm{GeV}^{-1}.
\end{eqnarray}
Both calculations based on Lattice QCD \cite{Becirevic:2009lq}
and QCD sum rules (QSR) \cite{Zhu:1996sr} predict that the coupling
constant for the radiative decay of the neutral charmed meson has a
positive sign while the coupling constant for the charged charmed
meson radiative decay is negative. In present work, we follow such a
convention and take $ g_{D^{*+} D^{+} \gamma} = -0.5$
$\mathrm{GeV}^{-1}$ and $\ g_{D^{*0}D^{0}\gamma} = 2.0$
$\mathrm{GeV}^{-1}$. For the coupling constant of $D_{s}^{*} D_{s}
\gamma$ interaction, the calculation from QSR gives $g_{D^{*}_s D_s
\gamma} =-0.3 \pm 0.1$ $\mathrm{GeV}^{-1}$ \cite{Zhu:1996sr}. In present
work, the central value is adopted.

According to the Lagrangian just listed above, we obtain the decay
amplitudes of $\chi_{c1}\to \gamma \rho^0$ corresponding to the diagrams in Fig. \ref{Fig:gpept-1},
\begin{eqnarray}
\mathcal{M}^{(a)}_C &=& (i)^3 \int \frac{d^4 q}{(2 \pi)^4} [i
g_{\chi_{c1}D D^{*}} \epsilon_{\chi_{c1}}^{\mu}] [-i g_{DDV}
\epsilon_{V}^{\nu}(i
q_{\nu} +i p_{1 \nu})]\nonumber\\
&&\times \Big[\frac{e}{4} g_{D^{*+}D^+ \gamma}
\varepsilon_{\theta \phi \lambda \kappa} \epsilon_{\gamma}^{\rho} (i
p_{4}^{\theta} g^{\phi}_{\rho}-i p_{4}^{\phi} g^{\theta}_{\rho}) (-i
p_{2}^{\lambda} g^{\tau \kappa} + i p_{2}^{\kappa} g^{\tau
\lambda})\Big]\nonumber\\
&&\times \frac{i}{p_{1}^2 -m_{D^+}^2} \frac{i (-g_{\tau \mu} +p_{2
\tau} p_{2 \mu}/m_{D^{*-}}^2)}{p_{2}^{2} -m_{D^{*-}}^2} \frac{i}{q^2
-m_{D^+}^2} \mathcal{F}^2(q^2),\label{11}
\end{eqnarray}
\begin{eqnarray}
\mathcal{M}^{(b)}_C &=& (i)^3 \int \frac{d^4 q}{(2 \pi)^4} [i
g_{\chi_{c1}D D^{*}} \epsilon_{\chi_{c1}}^{\mu}] [-2 f_{D^{*}DV}
\varepsilon_{\theta \nu \alpha \beta} \epsilon_{V}^{\nu}(i
p_{3}^{\theta}) (iq^{\alpha} +i p_{1}^{\alpha})]\nonumber\\
&&\times [-e \epsilon_{\gamma}^{\rho} (g_{\tau \phi} (p_{2
\rho}-q_{\rho})-g_{\rho \phi} q_{\tau} +g_{\rho \tau} p_{2 \phi})]
\frac{i}{p_{1}^2 -m_{D^+}^2} \frac{i (-g^{\tau \mu} +p_{2}^{\tau}
p_{2}^{\mu}/m_{D^{*-}}^2)}{p_{2}^{2} -m_{D^{*-}}^2} \nonumber\\
&&\times\frac{i (-g^{\beta \phi} +q^{\beta}
q^{\phi}/m_{D^{*}}^2)}{q^2 -m_{D^{*+}}^2} \mathcal{F}^2(q^2),
\end{eqnarray}
\begin{eqnarray}
\mathcal{M}^{(c)}_C &=& (i)^3 \int \frac{d^4 q}{(2 \pi)^4} [i
g_{\chi_{c1}D D^{*}} \epsilon_{\chi_{c1}}^{\mu}] [ie
\epsilon_{\gamma}^{\nu} (-i p_{1 \nu} -iq_{\nu})] [-2 f_{D^{*}DV}
\varepsilon_{\theta \rho \alpha \beta} \epsilon_{V}^{\rho} (i
p_{4}^{\theta}) (-i q^{\alpha}
+ip_{2}^{\alpha}) ]\nonumber\\
&&\times \frac{i}{p_{1}^2 -m_{D^+}^2} \frac{i (-g_{\mu}^{\beta} +p_{2
\mu} p_{2}^{\beta}/m_{D^{*-}}^2)}{p_{2}^2-m_{D^{*-}}^2}
\frac{i}{q^2-m_{D^+}^2} \mathcal{F}^2(q^2),
\end{eqnarray}
\begin{eqnarray}
\mathcal{M}^{(d)}_C &=& (i)^3 \int \frac{d^4 q}{(2 \pi)^4} [i
g_{\chi_{c1}D D^{*}} \epsilon_{\chi_{c1}}^{\mu}] [ \frac{e}{4}
g_{D^{*+}D^+\gamma} \varepsilon^{\theta \phi \alpha \beta}
\epsilon_{\gamma}^{\nu} (i p_{3 \theta} g_{\phi \nu}-ip_{3 \phi}
g_{\theta \nu}) (i q_{\alpha}
g_{\tau \beta} -i q_{\beta} g_{\tau \alpha})]\nonumber\\
&&\times[i g_{D^* D^* V} \epsilon_{V}^{\rho} (-i q_{\rho} +i p_{2
\rho})g_{\kappa \lambda} +4i f_{D^*D^*V} \epsilon^{\rho} (i p_{4
\kappa} g_{\rho \lambda} -i p_{4 \lambda} g_{\kappa
\rho})]\nonumber\\
&&\times \frac{i}{p_{1}^2 -m_{D^+}^2} \frac{i (-g_{\mu}^{ \kappa}
+p_{2 \mu} p_{2}^{\kappa}/m_{D^{*-}}^2)}{p_{2}^2-m_{D^{*-}}^2}
\frac{i(-g^{\lambda \tau} +q^{\lambda}
q^{\tau}/m_{D^{*+}}^2)}{q^2-m_{D^{*+}}^2} \mathcal{F}^2(q^2),\label{44}
\end{eqnarray}
which are resulted from the charge intermediate charmed mesons. Thus, the total decay amplitude
of $\chi_{c1}\to D\bar{D}^*+h.c.\to\gamma\rho^0$ is
\begin{eqnarray}
\mathcal{M}(\chi_{c1}\to D\bar{D}^*+h.c.\to\gamma\rho^0)=[\mathcal{M}^{(a)}_C+\mathcal{M}^{(b)}_C
+\mathcal{M}^{(c)}_C+\mathcal{M}^{(d)}_C]+[\mathcal{M}^{(a)}_N+\mathcal{M}^{(d)}_N],
\end{eqnarray}
where subscripts $C$ and $N$ denote the corresponding amplitudes being from charge charmed meson loop and
neutral charmed meson loop, respectively. $\mathcal{M}^{(a)}_N$ and $\mathcal{M}^{(d)}_N$ is obtained by
amplitudes $\mathcal{M}^{(a)}_C$ and $\mathcal{M}^{(d)}_C$ with the replacements of the mass and coupling constants, i.e., $g_{D^{*+}D^+\gamma}\to g_{D^{*0}D^0}\gamma$, $m_{D^{(*)+}}\to m_{D^{(*)0}}$ and  $m_{D^{(*)-}}\to m_{\bar{D}^{(*)0}}$. In Eqs. (\ref{11})-(\ref{44}), the form factor $\mathcal{F}(q^2)$ is introduced to depict the inner structure of the interaction vertex of the exchanged charmed
meson and the intermediated state. As what we have done in Ref.
\cite{Chen:2010rd}, a dipole form of the form factor is employed
\begin{eqnarray}
\mathcal{F}(q^2)=\left(\frac{\Lambda^2-m^2}{\Lambda^2-q^2}\right)^2.
\label{Eq:FFs}
\end{eqnarray}
Furthermore, the form factors with the pole form also play the role to
make the ultraviolet divergence disappear, in analog to the cutoffs
in the Pauli-Villas renormalization scheme. Here, the cutoff
$\Lambda$ can be parameterized as $\Lambda= m+ \alpha \Lambda_{QCD}$
with $\Lambda_{QCD}=220$ MeV and $m$ is the mass of the exchanged
meson in Fig. \ref{Fig:gpept-1} \cite{Cheng:2004ru}. We emphasize that a dipole form factor is introduced in the numerical calculation of this work, which was applied to the calculation of $\chi_{c1}\to VV$ $(V=\rho,\,\omega,\,\phi)$ in Ref. \cite{Chen:2010rd}.

The decay amplitude of $\chi_{c1}\to D\bar{D}^*+h.c.\to \gamma \omega$ is similar to that of $\chi_{c1}\to D\bar{D}^*+h.c.\to \gamma \rho^0$ besides multiplying an extra factor -1 since the $DD^*V$ vertex would have a different sign for $DD^*\omega$ and $DD^*\rho$ due to $SU(3)$ flavor symmetry. In addition, we need to replace the mass of $\rho$ with those of $\omega$. For $\chi_{c1}\to D_s\bar{D}_s^*+h.c.\to \gamma \phi$, the corresponding hadronic loops are composed of $D_{s}^{(*)+}$ and $D^{(*)-}_s$ charmed -strange mesons. The total decay amplitude is
\begin{eqnarray}
\mathcal{M}(\chi_{c1}\to D_s\bar{D}_s^*+h.c.\to\gamma\phi)=[\mathcal{M}^{(a)}_C+\mathcal{M}^{(b)}_C
+\mathcal{M}^{(c)}_C+\mathcal{M}^{(d)}_C]\bigg|_{m_{D^{(*)\pm}}\to m_{D_s^{(*)\pm}}}^{g_{D^{(*)+}D^{+}\gamma}\to g_{D_s^{(*)}D_s\gamma}}
\end{eqnarray}
with the replacements of the corresponding masses and coupling constants.

\section{Numerical Result}\label{sec3}

Before performing the numerical calculations, we need to introduce
the coupling constant relevant to the effective Lagrangian listed in
the previous section. The coupling constants for $\chi_{c1}
\mathcal{D} \mathcal{D}^{*}$ and $\mathcal{D} \mathcal{D}
\mathcal{V}$ interactions are listed in Table. \ref{Tab:parameter}.

\renewcommand{\arraystretch}{1.6}
\begin{table}[htb]
\centering
\begin{tabular}{c|cc|c|ccc}
\toprule[1pt]
 Coupling constant & Expression &Value &Coupling constant & Expression&Value\\\midrule[1pt]
  $g_{\chi_{c1 DD^{*}}}$ & $2\sqrt{2}g_1 \sqrt{m_D m_{D^{*}} m_{\chi_{c1}}}$ & $-21.44$ GeV &
  $g_{\chi_{c1 D_sD_s^{*}}}$ & $2\sqrt{2}g_1 \sqrt{m_{D_s} m_{D_s^{*}} m_{\chi_{c1}}}$ & $-22.60$ GeV \\
  $g_{\mathcal{D} \mathcal{D} \mathcal{V}}$ & $\beta g_{V}/\sqrt{2}$ & $3.71$ &
  $g_{\mathcal{D}^{*} \mathcal{D}^{*} \mathcal{V}}$ & $\beta g_{V}/\sqrt{2}$ & $3.71$\\
  $f_{\mathcal{D}^{*} \mathcal{DV}}$ & $\lambda g_{V}/\sqrt{2}$ & $2.31$ GeV$^{-1}$&
  $f_{D^{*} D^{*} \mathcal{V}}$ & $\lambda m_{D^{*}} g_{V}/\sqrt{2}$ & $4.64$\\
  $f_{D_s^{*} D_{s}^{*}\mathcal{V}}$& $\lambda m_{D_s^{*}} g_{V}/\sqrt{2}$
  & 4.88& & &\\
  \bottomrule[1pt]
\end{tabular}
\caption{The coupling constants relevant to the calculation of
$\chi_{c1}\to \gamma V$. Here, $g_1$ is related to the $\chi_{c0}$
decay constant $f_{\chi_{c0}}$ via relation
$g_1=-\sqrt{\frac{m_{\chi_{c0}}}{3}} \frac{1}{f_{\chi_{c0}}}$ with
$f_{\chi_{c0}} \simeq 0.51$ GeV \cite{Colangelo:2002bq}. Other
parameters include $g_V=m_\rho/f_\pi$, $m_\rho=0.77$ MeV,
$\beta=0.9$, $\lambda=0.56$ GeV$^{-1}$, $g=0.59$ and $f_\pi=132$ MeV
\cite{Casalbuoni:1996pg, Oh:2000qr, Isola:2003fh, Cheng:2004ru}.
\label{Tab:parameter}}
\end{table}

\begin{center}
\begin{figure}[htb]
\scalebox{0.55}{\includegraphics{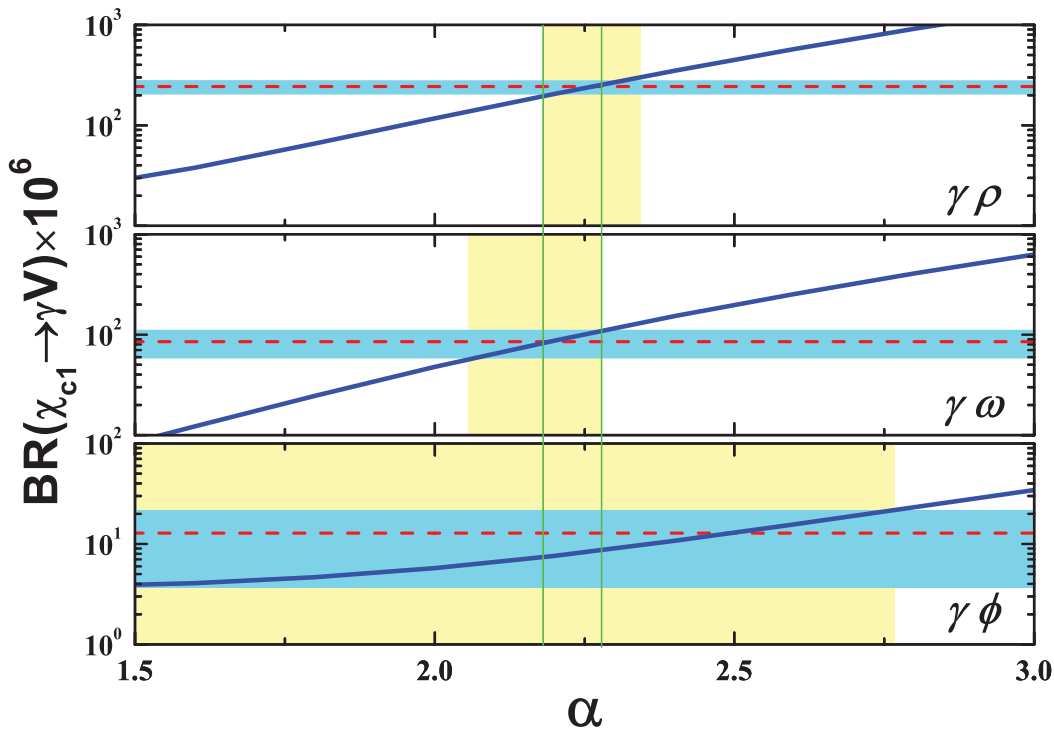}}%
\caption{(color online) The branching ratios of $\chi_{c1} \to
\gamma \rho^0,\ \gamma \omega,\ \gamma \phi$ dependent on the
parameter $\alpha$. The red dashed-lines with the blue bands are
the experimental measurement \cite{Bennett:2008aj}. The blue solid
lines correspond to the theoretical calculations including the
hadronic loop effect and the PQCD calculation \cite{Gao:2006bc}. The
vertical yellow bands in the sub-figure denote the overlap of our
results with the corresponding experimental measurement. With the
same $\alpha$ range sandwiched between two green vertical solid lines, the
obtained branching ratios of $\chi_{c1} \to \gamma \rho^0,\, \gamma
\omega,\, \gamma \phi$ in this work are consistent with the
experimental data. \label{Fig:radiative}}
\end{figure}
\end{center}

With the above preparation, the radiative decays of $\chi_{c1}$ to a
light vector meson are estimated. As a free parameter, $\alpha$ is
introduced in the cutoff $\Lambda$ of the form factors, which is
usually dependent on the particular process and taken to be of the
order of unity. In Fig. \ref{Fig:radiative}, we present the
branching ratios of $\chi_{c1} \to \gamma \rho^{0},\ \gamma \omega,\
\gamma \phi$ dependent on the parameter $\alpha$. For comparing with
the experimental data \cite{Bennett:2008aj}, the theoretical result
includes the hadronic loop contribution obtained in this work and the
PQCD estimation in Ref. \cite{Gao:2006bc}.

As shown in Fig. \ref{Fig:radiative}, there exists overlap
between the numerical result obtained in this work and the
experimental data announced by CLEO. The corresponding $\alpha$
ranges for $\chi_{c1} \to \gamma \rho^0,\, \gamma\omega,\, \gamma
\phi$ are $2.18 < \alpha <2.35$, $2.06 < \alpha <2.28$ and $1.16 <
\alpha <2.77$, respectively, which are in the reasonable parameter
space. Especially, we need to emphasize that there exists a common
$\alpha$ range $2.18 < \alpha <2.28$ for $\chi_{c1} \to \gamma
\rho^{0},\ \gamma \omega,\ \gamma \phi$ radiative decays.

\section{Discussion and Conclusion}\label{sec4}

As an important non-perturbative QCD effect, the hadronic loop
mechanism was proposed in studying $J/\psi$ and $\psi(3770)$ decays
\cite{Lipkin:1986av,Liu:2006dq,Liu:2009dr}. By using this mechanism,
the OZI suppressed processes $\chi_{c1} \to VV$ with $VV=\omega
\omega,\ \phi \phi$ and the double-OZI process  $\chi_{c1}\to \omega
\phi$ announced by BES-III were explained well in the recent work
\cite{Chen:2010rd,Liu:2010rd}.

The CLEO Collaboration announced the experimental results of
$\chi_{c1}\to \gamma V$ in 2008 \cite{Bennett:2008aj}, which are an
order magnitude larger than the corresponding theoretical
estimations calculated by PQCD \cite{Gao:2006bc}. To search for
the source of the discrepancy between the PQCD calculation
\cite{Gao:2006bc} and the CLEO data \cite{Bennett:2008aj} for
$\chi_{c1}\to \gamma V$, in this work we propose the hadronic loop
contribution to $\chi_{c1}\to \gamma V$. Under the hadronic loop
mechanism, $\chi_{c1}\to \gamma V$ is similar to $\chi_{c1}\to V V$
process, both of which occur via the intermediate $D\bar{D}^*$ just
shown in Fig. \ref{Fig:gpept-1}. To some extent, the study of this
work can be as a test for the hadronic loop effect, which was
applied to explain $\chi_{c1} \to VV$ decays
\cite{Chen:2010rd,Liu:2010rd}.

Our numerical result of $\chi_{c1}\to \gamma V$ indicates that the
theoretical result including the hadronic loop contribution and the
result in PQCD calculation for $\chi_{c1}\to \gamma V$ can reach up
to the experimental data of $\chi_{c1}\to \gamma V$. Thus,
non-perturbative QCD effect, i.e., hadronic loop mechanism, can be
as the underlying source to alleviate the difference between the
PQCD calculation and the CLEO data of $\chi_{c1}\to \gamma V$. As
indicated in Refs. \cite{Chen:2010rd,Liu:2010rd}, the hadronic loop
effect also plays an important role to $\chi_{c1}\to V V$. Thus, the
success of explaining $\chi_{c1}\to VV$ and $\chi_{c1}\to \gamma V$
under the hadronic loop mechanism not only tests the model itself, but
also shows that the non-perturbative effect on $\chi_{c1}$ is
important. Further experimental and theoretical studies of $\chi_{cJ}$ decay
are encouraged.

\section*{Acknowledgements}

We would like to thank Rong-Gang Ping from the BES Collaboration and V.~E.~Lyubovitskij for the discussion.
This project is supported by the National Natural Science Foundation
of China (NSFC) under Contracts No. 10705001, No. 10775148, No.
10975146; CAS Grant No.
KJCX3-SYW-N2; the Foundation for the Author of National Excellent
Doctoral Dissertation of P.R. China (FANEDD) under Contracts No.
200924; the Doctoral Program Foundation of Institutions of Higher
Education of P.R. China under Grant No. 20090211120029; the Program
for New Century Excellent Talents in University (NCET) by Ministry
of Education of P.R. China under Grant No. NCET-10-0442; the Fundamental Research Funds for the Central Universities.


\begin{thebibliography}{98}

\bibitem{Li:2008ey}
  X.~Q.~Li, X.~Liu and Z.~T.~Wei,
  Front.\ Phys.\ China {\bf 4}, 49 (2009)
  [arXiv:0808.2587 [hep-ph]].

\bibitem{Amsler:2008zzb}
  C.~Amsler {\it et al.}  [Particle Data Group],
  Phys.\ Lett.\  B {\bf 667}, 1 (2008).

\bibitem{zhangJ}J. Zhang {\it et al}. (BES-III Collaboration), in
Hadron 2009 Conference, Floarida State University.

\bibitem{Lipkin:1986av}
  H.~J.~Lipkin,
  Phys.\ Lett.\  B {\bf 179}, 278 (1986).

\bibitem{Liu:2006dq}
  X.~Liu, X.~Q.~Zeng and X.~Q.~Li,
  Phys.\ Rev.\  D {\bf 74}, 074003 (2006)
  [arXiv:hep-ph/0606191].

\bibitem{Li:2007xr}
  G.~Li and Q.~Zhao,
  Phys.\ Lett.\  B {\bf 670}, 55 (2008)
  [arXiv:0709.4639 [hep-ph]].

\bibitem{Liu:2009dr}
  X.~Liu, B.~Zhang and X.~Q.~Li,
  Phys.\ Lett.\  B {\bf 675}, 441 (2009)
  [arXiv:0902.0480 [hep-ph]].

\bibitem{Liu:2009iw}
  X.~Liu,
  Phys.\ Lett.\  B {\bf 680}, 137 (2009)
  [arXiv:0904.0136 [hep-ph]].

\bibitem{He:2006is}
  X.~G.~He, X.~Q.~Li, X.~Liu and X.~Q.~Zeng,
  Eur.\ Phys.\ J.\  C {\bf 51}, 883 (2007)
  [arXiv:hep-ph/0606015].

\bibitem{Liu:2007qi}
  X.~Liu, B.~Zhang, L.~L.~Shen and S.~L.~Zhu,
  Phys.\ Rev.\  D {\bf 75}, 074017 (2007)
  [arXiv:hep-ph/0701022].

\bibitem{Liu:2007fe}
  X.~Liu and B.~Zhang,
  Eur.\ Phys.\ J.\  C {\bf 54}, 253 (2008)
  [arXiv:0711.3813 [hep-ph]].

\bibitem{Liu:2007ez}
  X.~Liu,
  Eur.\ Phys.\ J.\  C {\bf 54}, 471 (2008)
  [arXiv:0708.4167 [hep-ph]].

\bibitem{Liu:2008yy}
  X.~Liu, B.~Zhang and S.~L.~Zhu,
  Phys.\ Rev.\  D {\bf 77}, 114021 (2008)
  [arXiv:0803.4270 [hep-ph]].

\bibitem{Dong:2009tg}
Y.~B.~Dong, A.~Faessler, T.~Gutsche, V.~E.~Lyubovitskij,
Phys. \ Rev.\ D {\bf 81}, 014006 (2010)
[arXiv: 0910.1204 [hep-ph]].

\bibitem{Faessler:2007gv}
  A.~Faessler, T.~Gutsche, V.~E.~Lyubovitskij and Y.~L.~Ma,
  Phys.\ Rev.\  D {\bf 76}, 014005 (2007)
  [arXiv:0705.0254 [hep-ph]];

\bibitem{Faessler:2007us}
  A.~Faessler, T.~Gutsche, V.~E.~Lyubovitskij and Y.~L.~Ma,
  Phys.\ Rev.\  D {\bf 76}, 114008 (2007)
  [arXiv:0709.3946 [hep-ph]];

\bibitem{Faessler:2008vc}
  A.~Faessler, T.~Gutsche, V.~E.~Lyubovitskij and Y.~L.~Ma,
  Phys.\ Rev.\  D {\bf 77}, 114013 (2008)
  [arXiv:0801.2232 [hep-ph]].

\bibitem{Dong:2008gb}
  Y.~B.~Dong, A.~Faessler, T.~Gutsche and V.~E.~Lyubovitskij,
  Phys.\ Rev.\  D {\bf 77}, 094013 (2008)
  [arXiv:0802.3610 [hep-ph]];

\bibitem{Dong:2009yp}
  Y.~B.~Dong, A.~Faessler, T.~Gutsche, S.~Kovalenko and V.~E.~Lyubovitskij,
  Phys.\ Rev.\  D {\bf 79}, 094013 (2009)
  [arXiv:0903.5416 [hep-ph]];

\bibitem{Branz:2009yt}
  T.~Branz, T.~Gutsche and V.~E.~Lyubovitskij,
  Phys.\ Rev.\  D {\bf 80}, 054019 (2009)
  [arXiv:0903.5424 [hep-ph]];

\bibitem{Dong:2010gu}
  Y.~Dong, A.~Faessler, T.~Gutsche and V.~E.~Lyubovitskij,
  Phys.\ Rev.\  D {\bf 81}, 074011 (2010)
  [arXiv:1002.0218 [hep-ph]].

\bibitem{Liu:2006df}
  X.~Liu, B.~Zhang and S.~L.~Zhu,
  Phys.\ Lett.\  B {\bf 645}, 185 (2007)
  [arXiv:hep-ph/0610278].

\bibitem{Chen:2010rd}
  D.Y. Chen, J. He, X.Q. Li and X. Liu,
  Phys.\ Rev.\ D {\bf 81} 074006 (2010).

\bibitem{Liu:2010rd}
  X.H. Liu and Q. Zhao,
  Phys.\ Rev.\ D {\bf 81} 014017 (2010).

\bibitem{Gao:2006bc}
  Y.~J.~Gao, Y.~J.~Zhang and K.~T.~Chao,
  Chin.\ Phys.\ Lett.\  {\bf 23}, 2376 (2006)
  [arXiv:hep-ph/0607278].

\bibitem{Bennett:2008aj}
  J.~V.~Bennett {\it et al.}  [CLEO Collaboration],
  Phys.\ Rev.\ Lett.\  {\bf 101}, 151801 (2008)
  [arXiv:0807.3718 [hep-ex]].

\bibitem{Colangelo:2003sa}
  P.~Colangelo, F.~De Fazio and T.~N.~Pham,
  Phys.\ Rev.\  D {\bf 69}, 054023 (2004)
  [arXiv:hep-ph/0310084].

\bibitem{Cheng:1992xi}
  H.~Y.~Cheng, C.~Y.~Cheung, G.~L.~Lin, Y.~C.~Lin, T.~M.~Yan and H.~L.~Yu,
  Phys.\ Rev.\  D {\bf 47}, 1030 (1993)
  [arXiv:hep-ph/9209262].

\bibitem{Yan:1992gz}
  T.~M.~Yan, H.~Y.~Cheng, C.~Y.~Cheung, G.~L.~Lin, Y.~C.~Lin and H.~L.~Yu,
  Phys.\ Rev.\  D {\bf 46}, 1148 (1992)
  [Erratum-ibid.\  D {\bf 55}, 5851 (1997)].

\bibitem{Wise:1992hn}
  M.~B.~Wise,
  Phys.\ Rev.\  D {\bf 45}, R2188 (1992).

\bibitem{Burdman:1992gh}
  G.~Burdman and J.~F.~Donoghue,
  Phys.\ Lett.\  B {\bf 280}, 287 (1992).

\bibitem{Casalbuoni:1996pg}
  R.~Casalbuoni, A.~Deandrea, N.~Di Bartolomeo, R.~Gatto, F.~Feruglio and G.~Nardulli,
  Phys.\ Rept.\  {\bf 281}, 145 (1997)
  [arXiv:hep-ph/9605342].

\bibitem{Dong:2009uf}
  Y.~Dong, A.~Faessler, T.~Gutsche and V.~E.~Lyubovitskij,
  arXiv:0909.0380 [hep-ph].

\bibitem{Colangelo:1994jc}
  P.~Colangelo, F.~De Fazio and G.~Nardulli,
  Phys.\ Lett.\  B {\bf 334}, 175 (1994)
  [arXiv:hep-ph/9406320].

\bibitem{Zhu:1996sr}
  S.~L.~Zhu, W.~Y.~Hwang and Z.~S.~Yang,
  Mod.\ Phys.\ Lett.\  A {\bf 12}, 3027 (1997)
  [arXiv:hep-ph/9610412].

\bibitem{Cheng:2004ru}
  H.~Y.~Cheng, C.~K.~Chua and A.~Soni,
  Phys.\ Rev.\  D {\bf 71}, 014030 (2005)
  [arXiv:hep-ph/0409317].

\bibitem{Colangelo:2002bq}
  P. Colangelo, F. De Fazio and T. N. Pham,
  Phys. Lett. B {\bf 542}, 71 (2002).

\bibitem{Oh:2000qr}
  Y.~S.~Oh, T.~Song and S.~H.~Lee,
  Phys.\ Rev.\  C {\bf 63}, 034901 (2001)
  [arXiv:nucl-th/0010064].

\bibitem{Isola:2003fh}
  C.~Isola, M.~Ladisa, G.~Nardulli and P.~Santorelli,
  Phys.\ Rev.\  D {\bf 68}, 114001 (2003)
  [arXiv:hep-ph/0307367].

\bibitem{Becirevic:2009lq}
D.~Becirevic and B.~Haas,
  arXiv:0903.2407 [hep-lat].




\end{thebibliography}
\end{document}